\documentclass[conference,10pt]{IEEEtran}
\usepackage{url}
\usepackage{amsmath,amssymb,amsfonts}
\usepackage{algorithmic}
\usepackage{graphicx}
\usepackage{textcomp}
\usepackage[hidelinks]{hyperref}
\usepackage{xcolor}

\usepackage[style=ieee, backend=biber]{biblatex}

\begin{document}

\title{Hardware Design and Security in the Era of Chiplets and LLMs}

\author{\IEEEauthorblockN{Johann Knechtel}
\IEEEauthorblockA{NYU Abu Dhabi\\
johann@nyu.edu}
\and
\IEEEauthorblockN{Ozgur Sinanoglu}
\IEEEauthorblockA{NYU Abu Dhabi\\
ozgursin@nyu.edu}
\and
\IEEEauthorblockN{Paul V. Gratz}
\IEEEauthorblockA{Texas A\&M University\\
pgratz@tamu.edu}
\and
\IEEEauthorblockN{Ramesh Karri}
\IEEEauthorblockA{NYU Tandon School of Engineering\\
rkarri@nyu.edu}
}

%\author{%
%\IEEEauthorblockN{Author 1}
%\IEEEauthorblockA{Affiliation 1\\
%email1@edu}
%\and
%\IEEEauthorblockN{Author 2}
%\IEEEauthorblockA{Affiliation 2\\
%email2@edu}
%\and
%\IEEEauthorblockN{Author 3}
%\IEEEauthorblockA{Affiliation 3\\
%email3@edu}
%\and
%\IEEEauthorblockN{Author 4}
%\IEEEauthorblockA{Affiliation 4\\
%email4@edu}
%}

\maketitle

\begin{abstract}
The semiconductor industry is undergoing a dual revolution: the shift toward
heterogeneous 2.5D chiplet systems and the integration of Large Language
Models (LLMs) into Electronic Design Automation (EDA) flows. While these
paradigms offer unprecedented benefits in yield, modularity, design
productivity, etc., they radically expand the hardware attack surface. This paper
provides a unified analysis of these frontiers,
	 %covering threats
	 ranging from
attacks on chiplet systems (including hardware stacks for LLM acceleration) across architectural, logical, and physical levels,
%level (interconnect snooping and cache-coherence exploits) and physical level (side-channel and fault injection attacks),
to various exploits against LLM-driven EDA pipelines.
%(backdoors, IP leakage, data/benchmark contamination).
To secure chiplet systems, we review a powerful defense approach that leverages 2.5D split
manufacturing and active interposers for physically isolated Root of Trust
(RoT) architectures.
%The resulting 2.5D RoT directly supports
%%handle the twofold threats of
%secure system-level integration of untrusted chiplets and integration of other security features like sensors.
%%attacks on chiplets hardware.
%We also discuss physical design benefits of such a 2.5D RoT.
%%showing significant improvements in power and signal integrity alongside a
%%reduction in total system footprint.
To secure LLM-driven EDA pipelines, we first identify native threats and then review state-of-the-art defense techniques.
Finally, we discuss how LLM systems can advance hardware security efforts
for modern systems, including chiplets.
\end{abstract}

% NOTES
% 
% 2.5D RoT for trustworthy LLM acceleration; add specific security risks for LLM accel on chiplets OR clearly map described threats to LLM accel
% LLMs for design, security policies, red teaming etc for chiplet systems

\section{Introduction}

%To overcome the CMOS scalability bottleneck,
The semiconductor industry is rapidly transitioning from monolithic Systems-on-Chips (SoCs) toward high-density 2.5D chiplet-based architectures
\cite{naffziger2021pioneering, apple2022m1ultra, nassif2022sapphire}.
This paradigm enables a modular ``plug-and-play'' ecosystem where third-party chiplets are manufactured separately and integrated onto a silicon interposer.
%\cite{kim2019architecture}.
However, the reliance on multi-vendor supply chains introduces significant trust gaps, as individual chiplets may originate from untrusted foundries or design houses, threatening the
%security and trustworthiness of
overall system's integrity \cite{nabeel2020, chacon2024}.
%, in order to utilize cache-coherent memory and other system-level resources,
These chiplets must share a common interconnect fabric and memory space, making
system-level communication a highly vulnerable surface \cite{chacon2024, suzano2024hardware}.
Such threats apply to any chiplet system, including heterogeneous accelerator architectures for deployment of large language models (LLMs)
\cite{10764574, peng2024chipletcloudbuildingai, chen2025chimechipletbasedheterogeneousnearmemory}.

Independently, LLMs for electronic design automation (EDA) tooling have significantly accelerated hardware-design workflows \cite{SurveyChipDesign, SurveyLLM, SurveyEDA}.
For example, transformer-based architectures automate hardware description language (HDL) coding and formal verification \cite{VeriGen, AutoChip}.
However, this automation also introduces threats for hardware design, including leakage of design IP, backdoor poisoning, and data/benchmark contamination \cite{VeriLeaky, RTLBreaker, VeriContaminated}.
Simultaneously, LLM frameworks become capable of assisting security measures, e.g.,
%implementing and verifying hardware security measures, e.g.,
%%For example, proposing orchestrated LLM agents, researchers
%orchestrated agents can advance red-teaming efforts for
%security assessment of logic locking and IP piracy \cite{LockForge,NetDeTox} and
automatic generation of security assertions \cite{SecurityAssertions}.

Protecting this modern and complex landscape of chiplet systems and wide-scale use of LLMs requires new solutions on multiple fronts.
For the hardware, a fundamental shift from chip-level security to system-level architectures built on physically isolated Roots of Trust is required.
For LLM operation, orchestrating agents within trusted boundaries for hardware, software, and data operation is required.
Crucially, there is a lack of cross-domain understanding bridging these paradigms.
While attackers are poised to exploit fuzzy boundaries when multi-vendor fabrics interface with sensitive acceleration hardware,
current research fails to leverage trustworthy LLMs to secure complex chiplet systems and vice versa.

Here, we review the state-of-the-art in chiplet security, focusing on interposer-based defenses.
We also unravel the dynamics of LLM-native EDA flows and their promises for hardware design, alongside related risks and emerging defenses.
Finally, we are charting a path toward synergetic and secure operation of chiplet systems and LLM pipelines.

\section{Background}

\subsection{Chiplet Systems}
%: A Plug-and-Play Ecosystem}

2.5D integration uses interposer to interconnect separately manufactured chiplets, reducing time-to-market and improving yield compared to monolithic SoCs \cite{vivet20, naffziger21}.\footnote{Aside from 2.5D integration, 3D IC
technologies like through-silicon via (TSV) stacked logic,
face-to-face bonding, and monolithic 3D integration have gained significant momentum as well.
See also \cite{knechtel2019} for such technologies in general and for related hardware-security aspects in particular.}
While current systems are mostly homogeneous and cache-coherent, the trend toward heterogeneity allows designers to integrate different commodity chiplets, each designed and manufactured for an
optimized process \cite{naffziger21, vivet20}.

Interposers are classified as passive (wiring only) or active (wiring and logic) \cite{jerger15}.
Active interposers with network-on-chip (NoC) infrastructure
%facilitate large-scale communication by
separate the
interposer's fabric from chiplet interconnects, enabling cross-optimized networking topologies
%, enhanced system design opportunities,
and improved testability
\cite{activeNoC_2018isca}.

\subsection{LLM-Driven Hardware Design: Challenges and Solutions}

\subsubsection{From Prompting to Optimization and Reasoning}
Early works often yielded flawed code, leading to feedback-driven agents like AutoChip~\cite{AutoChip} that autonomously correct compiler errors. \cite{MCTS_RTL} treats
register-transfer level (RTL) code
generation as a state-space search with backtracking. Frameworks such as VeriThoughts~\cite{VeriThoughts} utilize reasoning models like DeepSeek-R1 to generate Chain-of-Thought traces, employing formal verification to mitigate hallucinations.

%\subsubsection{Representation Bottlenecks and Alternatives}
%Intermediate representation significantly impacts performance; while Verilog achieves high pass rates due to relative data abundance, HLS-C suffers from an ``accessibility-competence paradox,'' where user-friendly
%IRs yield the worst results due to data scarcity~\cite{RepresentationBottleneck}. Beyond text, RTL++~\cite{RTLpp} enhances structural awareness by encoding RTL as graph representations (CFG/DFG). Alternatively,
%Veritas~\cite{Veritas} fine-tunes models to output conjunctive normal form (CNF) clauses, ensuring correct-by-construction synthesis via propositional logic.
%
%\subsubsection{High-Level Synthesis}
%Bridging the software-to-hardware gap, C2HLSC~\cite{C2HLSC} uses LLMs to refactor standard C into synthesizable HLS-C. For neural network accelerators, retrieval-augmented generation (RAG) and reasoning and acting
%(ReAct) pipelines facilitate automated HLS pragma injection (e.g., pipelining) based on synthesis reports~\cite{NNFPGA}. Furthermore, frontier reasoning models enable autonomous, constraint-aware design-space exploration (DSE)~\cite{ReasoningHLS}.

\subsubsection{Representation and Synthesis Bottlenecks}
Intermediate representations constrain LLM coding performance. While sufficient Verilog data allows for high pass rates, High-Level Synthesis (HLS-C) suffers from an ``accessibility-competence
paradox'' due to data scarcity~\cite{RepresentationBottleneck}.
To bridge this software-to-hardware gap, frameworks like C2HLSC~\cite{C2HLSC} utilize LLMs to refactor standard C into synthesizable HLS-C, while automated pipelines inject pragmas based on synthesis
reports~\cite{NNFPGA}.
Beyond text, enhancing structural awareness via graph representations (RTL++~\cite{RTLpp}) or fine-tuning models to output conjunctive normal-form clauses (Veritas~\cite{Veritas}) ensures correct-by-construction synthesis through propositional logic.

\subsubsection{Assertions and Testbenches}
LLMs can generate SystemVerilog Assertions (SVA)~\cite{SecurityAssertions}, though grounding them in RTL semantics remains challenging~\cite{viswambharan2026knowledgegraphsmissinglink}.
Deterministic decoding and specialized retrieval-augmented generation (RAG) frameworks~\cite{NL2SVA_RAG} improve SVA accuracy.
Coverage-driven agents can iteratively expand testbenches to achieve near-total transition coverage for complex state machines~\cite{FSM_Testbench}.

\subsubsection{Model Orchestration, Configuration, and Evaluation}
VeriDispatcher~\cite{VeriDispatcher} reduces API costs by 40\% by routing tasks to models based on predicted difficulty. Evaluation via the Hardware Quality Index~\cite{HQI_Eval} has identified failure modes like
complexity timeouts in frontier models. Crucially, inference-time configuration can impact pass rates by over 25\%; since optimal hyperparameters do not transfer across benchmarks, task-aware calibration is more vital than model size~\cite{ConfigOverSelection}.

\section{Threats for Chiplet Systems}

Multi-vendor chiplet systems introduce significant risks once chiplet design and fabrication are being outsourced, compromising the trust of system integration
\cite{nabeel2020, chacon2024}. Traditional security primitives, such as ARM TrustZone or Intel SGX, fail to extend protection across chiplet boundaries, leaving the system vulnerable
%to hardware and/or software attacks operating across chiplets
\cite{nabeel2020}.

\subsection{Architectural and Logical Attacks}

The shared interconnect and memory space form the primary vector for architectural exploits~\cite{chacon2024, suzano2024hardware}.
%Threat modeling must focus on malicious chiplets that may exploit system-level communication behaviors.
Similar to NoC vulnerabilities in concept~\cite{charles2021nocattksurvey}, yet different in execution,
2.5D systems are prone to inter-chiplet communication attacks like snooping and spoofing~\cite{nabeel2020, chacon2024}.
%snooping and spoofing, modification and diversion, and man-in-the-middle.
%\begin{itemize}
%\item {Snooping \& Spoofing:} Observing broadcast traffic for illicit data gathering or forging sender IDs to illegally request services.
%\item {Modification \& Diversion:} Tampering with payload data or rerouting packets by manipulating destination fields.
%\item {Man-in-the-Middle:} Hijacking and manipulating active transactions between legitimate chiplets.
%\end{itemize}
%As indicated,
The underlying threat for these attacks are malicious chiplets undermining the security of other chiplets when integrated into the same 2.5D system.
%without dedicated measures.
%%like a dedicated RoT architecture.

%\subsection{Logical Attacks: Cache Coherence and Microarchitecture}

In systems with cache coherence (CC), memory consistency is maintained across chiplets through low-level communication.
However, CC protocols operate transparently to operating system (OS)-level permissions, allowing Trojans in CC controllers to bypass page-table protections and
manipulate OS-restricted memory regions
by injecting malicious, unverified coherence responses \cite{chacon2024}.
Protocols like MOESI Hammer \cite{conway2010moesihammer}, which rely on broadcast traffic, are susceptible to multi-stage {Forging Attacks} \cite{chacon2024}.
%\footnote{(1) The Trojan snoops a target address; (2) it requests exclusive ownership from the memory directory; (3) it intercepts and hoards invalidation ACKs; and (4) it injects a malicious writeback to overwrite main memory.}
This exploit requires no address space ownership, leaving the global CC mechanism and the victim chiplet unaware of the compromise.
Shared CC fabrics also expose chiplets to microarchitectural threats: attackers can infer memory access patterns through NoC contention and latency without issuing unauthorized requests \cite{chacon2022}.
Trojans may also leverage speculative execution by issuing out-of-bounds prefetch requests that utilize CC protocol state changes to reveal information about a victim chiplet's working set \cite{chacon2024}.

While applicable to any 2.5D system, such threats are particularly concerning for heterogeneous LLM hardware stacks, with general-purpose CPUs, dedicated accelerator cores, and shared memories arranged across different chiplets \cite{10764574, peng2024chipletcloudbuildingai, chen2025chimechipletbasedheterogeneousnearmemory}.
For example, LLM inference demands large-scale memory transfers for weights and activations; the resulting multi-chiplet traffic exhibits highly predictable patterns.
Malicious third-party chiplets could exploit NoC and CC vulnerabilities discussed above, aiming to reconstruct model architectures or steal proprietary weights.
The current literature scrutinizes LLM stacks on monolithic SoCs
\cite{wu2026cache, 309838};
structural vulnerabilities of multi-vendor chiplet systems executing distributed acceleration remain a critical and unaddressed security gap.

\subsection{Physical Attacks}

%In addition to architectural and logical attacks,
%Physical attacks constitute another class of severe threats as follows.
Side-channel attacks (SCAs) exploit physical leakage such as power consumption, timing variations, and speculative cache states to infer sensitive data \cite{zhou05, lipp18}.
Fault-injection attacks (FIAs) use techniques like laser and electromagnetic pulses \cite{cui17} to fix/glitch specific signals, e.g., to assist SCAs or to disrupt
security features.
There are also indirect FIA methods
like memory hammering (RowHammer) or exploiting dynamic voltage and frequency scaling (DVFS) \cite{vanderVeen16}.
Using failure-analysis equipment like invasive ion-beam milling or semi-invasive electro-optical probing, adversaries can execute read-out attacks on logic and memory \cite{wang17_probing}.

Recent studies have demonstrated physical attacks also on LLM systems
\cite{hossain2026transformermemorycorruptedinvestigating, das2026}.
While most if not all of these studies are limited to LLM software stacks operating on regular SoCs, future attacks on chiplet-based LLM systems are equally realistic and concerning.

\section{Securing Chiplet Systems}
%: Physical and Architectural Foundations}

Most 2.5D Root of Trust (RoT) proposals utilize a centralized chiplet for authentication and security management \cite{liu2023securing}.
However, untrusted manufacturing can compromise that RoT, and its limited visibility allows malicious chiplets or NoC-level Trojans to bypass defenses \cite{charles2021nocattksurvey}.
Furthermore, traditional NoC security techniques \cite{charles2021nocattksurvey} assume a single trusted designer;
in multi-vendor systems, the interconnect itself may be compromised, necessitating security monitors that are physically isolated from the NoC fabric.

\subsection{2.5D RoT:
	%Split Manufacturing and Active Interposers as
		Physical and Architectural Foundations}
%\subsection{Split Manufacturing and Active Interposers for 2.5D RoT}

Split manufacturing, traditionally proposed for 2D and 3D ICs \cite{9344587, patnaik2019modern},
can also secure 2.5D integration, e.g., against IP piracy and Trojans \cite{xie2017security, knechtel2019}.
By concealing system-level routing within interposers, designers hide the full netlist from chiplet foundries, obscuring IP context and potential attack targets \cite{xie2017security, knechtel2019, patnaik2019modern}.
Importantly, this concept can also be used for physical separation of trusted security features and untrusted commodity chiplets as described next.

%Migrating to active interposers further enhances security by enabling logic splitting between trusted and untrusted facilities \cite{knechtel2019}.
%While passive interposers provide only wiring,
Active interposers can incorporate sensitive logic such as secured NoC routers,
%, repeaters, and voltage regulators,
enabling them to function as a physically isolated 2.5D RoT \cite{park2020, nabeel2020}.
Using mature nodes, active interposers can be fabricated in trusted on-shore facilities \cite{park2020}.
The 2.5D RoT enforces strict physical
separation: untrusted commodity chiplets must rely entirely on the interposer's fabric for communication, rendering attacks like spoofing impossible, as network interfaces reside within the trusted
interposer logic \cite{chacon2024, nabeel2020}.
Other attacks can be mitigated by runtime monitoring as discussed next.

\subsection{Runtime Monitors: Microarchitecture and Operation}

A 2.5D RoT built from an active interposer can embed Transaction Monitors (TRANSMONs) as hardware-level shims between untrusted chiplets and the interposer's NoC fabric \cite{nabeel2020}.
Such TRANSMONs provide multi-layer protection:
%an Address Protection Unit (APU)
	they enforce access control via core ID and policy checks, and
	%and Policy Register Spaces (PRSs), utilizing a Slave Access Filter (SAF) to drop illicit requests;
%concurrently, a Data Protection Unit (DPU) employs
data-masking hinders ``shadow writes'' against leakage of cryptographic assets etc.
Managed by an on-interposer secure core, these units implement a
deny-by-default stance.
For memory integrity, TRANSMONs facilitate parallel ECC/CRC verification by offloading metadata to trusted memory chiplets, neutralizing read-latency overheads \cite{nabeel2020}.

Crucially, as TRANSMONs are CC-agnostic, Coherence Message Checkers (CMCs) can be integrated into the NoC ingress to prevent protocol-level exploits \cite{chacon2024}.
%CMCs utilize a Packet Checker/Modifier (PCM) to
They validate coherence flits against secure OS-managed permissions in two variants: CMC-1 sits at the chiplet-NoC boundary to intercept malicious incoming flits, while CMC-2 is integrated with memory controllers to neutralize snooping.
In broadcast-heavy protocols like MOESI Hammer, CMC-2 dynamically transforms broadcasts into targeted unicasts or NACKs, effectively masking transaction existence from unauthorized cores with minimal pipeline latency (1--2 stages) \cite{chacon2024}.

Beyond mitigating architectural and logical attacks, the 2.5D RoT interposer fabric can provide trusted integration of sensors and other hardware security features against physical attacks.
This is especially promising for LLM hardware stacks that are composed of heterogeneous chiplets from multiple vendors, where individual chiplets may offer limited built-in security features
\cite{10764574, peng2024chipletcloudbuildingai, chen2025chimechipletbasedheterogeneousnearmemory}.
Furthermore, 3D ICs can provide comprehensive shielding structures across metal layers and TSVs, mitigating a wide range of physical attacks directly by construction \cite{knechtel2019, 9424027}.

\subsection{Physical Design and Performance Impact}

%The shift to utilizing active interposers requires moving from package-centric EDA tools to standard layout synthesis but with capabilities for handling multiple chiplets.
%A co-design flow implements chiplets in advanced nodes while synthesizing the interposer in mature nodes
%%; floorplanning projects chiplet microbumps as interposer pins and manages TSV/C4 placements for system-wide connectivity
%\cite{park2020}.

Active interposers fundamentally improve power and signal integrity \cite{park2020}:
embedding integrated voltage regulators (IVRs) helps to minimize the distance to power-demanding logic, reducing the maximum IR-drop by 73.7\%, and embedded repeaters mitigate signal distortion by eliminating long, unbuffered wires \cite{park2020}.
An active 2.5D design can reduce the total system silicon footprint by 18.5\% because the NoC and IVRs are removed from the surface-level chiplets \cite{park2020}.
At the same time, the interposer maintains a low utilization rate of 2.68\%, avoiding concerns for yield loss of interposers \cite{park2020}.

The performance impact for CMC-driven runtime monitoring is $\approx$4\% on average, while total system power is reduced by 3.2\% compared to unsecured baselines \cite{nabeel2020, chacon2024, park2020}.

\section{LLM-Driven EDA Flows: Threats and Defenses}

As design houses increasingly adopt LLM-driven EDA workflows, establishing trust in model training, operation, and EDA tool integration becomes critical.
Prominent threats and state-of-the-art defenses are discussed next.

\subsubsection{Backdoor Attacks}

RTL-Breaker~\cite{RTLBreaker} demonstrates that fine-tuning on public RTL repositories exposes models to data poisoning.
Attackers can embed rare keywords as activation triggers that cause the backdoored model to generate Trojans or sub-optimal circuitry while maintaining valid syntax,
allowing these modifications to evade standard functional checks.

SafeTune~\cite{rezakhani2026safetunemitigatingdatapoisoning} addresses
poisoning through offline dataset sanitization with an online inference guard,
%Before fine-tuning, the framework filters the corpus using a semantic prompt selector alongside a structural RTL filter that identifies Trojan payloads within DFGs.
%Inference-time paraphrasing tackles residual trigger patterns,
%effectively disrupting backdoor activation and
lowering the attack success rate down to 33\%.
Semantic Consensus Decoding (SCD)~\cite{yang2026semanticconsensusdecodingbackdoor}
%defends this threat by exploiting the
exploits the
locality bias, assuming adversaries embed triggers within non-functional modifiers.
%rather than functional hardware specifications.
At inference time,
   %a trigger-agnostic extractor isolates
   core functional requirements are extracted from the user prompt;
%SCD then computes the root-mean-square divergence between the logit distributions of the full specification and the functionally extracted counterpart.
%When a semantic trigger activates a backdoor---which is detected through
significant divergences in full vs extracted specification trigger
the framework
%adjusts an adaptive weight to
%suppress the malicious components and
to fall back to clean functional generation, reducing the observed attack rate to near 0\%.

\subsubsection{Data Contamination}

This leaks benchmark test sets into training data, artificially inflating evaluation scores through memorization.
VeriContaminated~\cite{VeriContaminated} utilizes Min-K\% probability and Contamination Detection via Distribution (CDD) metrics to reveal near-100\% contamination of standard benchmarks like VerilogEval across
recent commercial models.

Mitigating this threat requires a shift to dynamic benchmarking~\cite{chen-etal-2025-benchmarking-large} or model sanitization~\cite{SALAD}.
%, though strict filtering thresholds inherently degrade coding fluency.
Dynamic benchmarking establishes time-variant testing corpora that shift along model timestamps and/or utilizes rule-/template-based problem generation to mitigate collision rates~\cite{chen-etal-2025-benchmarking-large}.
Alternatively, when data exposure has already occurred, model sanitization can be pursued post-hoc via machine unlearning frameworks like SALAD~\cite{SALAD}.
%Rather than necessitating a complete, computationally prohibitive retraining lifecycle,
This framework leverages tailored loss primitives like gradient difference or Negative Preference Optimization (SimNPO) to sever the model's memory of contaminated test data (and malicious backdoors and sensitive IP)
while retaining general RTL coding fidelity~\cite{SALAD}.

\subsubsection{Safety Misalignment}

Prompt injection (PI) poses a fundamental risk to LLM systems~\cite{info17010054}.
A related concern is that established safety alignment lacks hardware-domain understanding.
HarmChip~\cite{HarmChip}, a recent safety benchmarking effort covering 16 domains and 120 threats,
revealed that keyword-sensitive guardrails indiscriminately block legitimate engineering
tasks, while adversaries can bypass the same guardrails using semantic disguise, e.g., by framing attacks as engineering change order (ECO) optimizations.

Defending against PI attacks demands structured runtime validation frameworks that treat untrusted instructions separately from system prompts~\cite{info17010054}.
%Rather than relying on simple keyword blacklists—which fail against sophisticated injection techniques like virtualization or adversarial multi-turn loops—robust defense implementations employ dual-core validation pipelines~\cite{info17010054}.
These frameworks combine embedding-based classification layers with strict structural parsing,
      %to encapsulate external user requests,
      neutralizing embedded trigger sequences before they interface with core execution layers.
To address the safety-alignment gap highlighted by HarmChip, frameworks must transcend general-purpose defense tuning by embedding domain-specific engineering logic directly into the model's safety
boundaries~\cite{HarmChip}.
By executing multi-task optimization using dual-purpose adversarial pairs, models can be calibrated via preference optimization algorithms (such as DPO and SimNPO) to evaluate the true underlying intent.
%rather than syntax.
%, successfully reducing jailbreak vulnerabilities while preserving unhindered hardware design utility~\cite{2604.17093v1 (2).pdf}.

\subsubsection{IP Leakage}

While promising for design quality, fine-tuning on in-house codes risks leaking the underlying IP.
VeriLeaky~\cite{VeriLeaky} demonstrated that, by providing only few structural hints like interface declarations, models can be coerced into flawlessly regenerating sensitive IP.

VeriLeaky also showed that logic locking can be applied before fine-tuning to reduce leakage, but it comes at the cost of training utility~\cite{VeriLeaky}.
As indicated, SALAD~\cite{SALAD} resolves this and other threats via machine unlearning.
CircuitGuard~\cite{11310859} proposes a dual-stream mechanism: pre-training sanitization and vocabulary masking.
%) to disrupt structural memorization.
%By utilizing abstract syntax tree (AST) parsers to identify proprietary design signatures and
Applying adaptive token-level noise masks during gradient descent optimization,
CircuitGuard mitigates both verbatim and behavioral replication of proprietary IP without modifications to the baseline architecture.
%As indicated, SALAD~\cite{SALAD} targets this challenge post-hoc via machine unlearning, leveraging algorithms like DPO and SimNPO to surgically strip the model's memory of targeted hardware modules.
%Multiple algorithms (e.g., SimNPO) are explored to scrub proprietary IP, backdoors, and contaminated benchmarks without full retraining,
%while preserving Verilog coding performance.

\section{LLMs for Hardware Security}

%Aside from advancing EDA flows in general,
	LLMs are also increasingly utilized to implement and verify hardware security across various settings.

\subsubsection{IP Protection}
Logic locking utilizes key-dependent gates to protect design IP.
GLLaMoR~\cite{GLLaMoR} speeds up traditional topological analysis by converting netlists to lock into adjacency lists for multi-hop graph reasoning in LLMs.
LockForge~\cite{LockForge} devises a multi-agent setup (Coder, Judge, Examiner) to parse academic PDFs and validate locking implementations via similarity scoring.
Similarly, the framework in~\cite{ghimire2026agentssecurehardwareevaluating} couples RAG with SAT-based iterative refinement for locking.
For IP redaction, ARIANNA~\cite{ARIANNA} automates FPGA design-space exploration (DSE) using module clustering and branch-and-bound algorithms, cutting area overhead by 3.3$\times$.

\subsubsection{SCA Mitigation and Cryptography Accelerators}
Traditional pre-silicon SCA assessment requires exhaustive design and power simulation runs.
NetlistWhisperer~\cite{NetlistWhisperer} uses an ensemble of fine-tuned LLMs to predict gate-level leakage bounds and to secure netlist implementations via Domain-Oriented Masking (DOM).
For Post-Quantum Cryptography (PQC), LLM4PQC~\cite{LLM4PQC} proposes an agentic workflow to isolate critical kernels, refactor dynamic arrays and other C constructs from reference codes (to resolve HLS obstacles), and
run DSE for FPGA implementation.
The framework has been extended to generate SCA-resilient accelerators~\cite{vts26_special_session_secure_PQC}.

\subsubsection{Trojan Detection}
Prior art utilizes graph neural networks (GNNs) operating on gate-level netlists directly or on graph representations of RTL~\cite{GNN_survey};
both views discard textual context from the original HDL, undermining detection accuracy.
In contrast, TrojanLoC~\cite{TrojanLoC}
%bypasses graph translation by processing raw Verilog directly.
Trained on the TrojanInS dataset, it uses
devises an RTL-adapted transformer to extract token- and line-level embeddings, achieving a 99\% F1-score for module-level detection and precise payload localization via lightweight classifiers.

\subsubsection{Red-Teaming}
%and Adversarial Evasion}
Evaluating security defenses requires adversarial red-teaming to uncover blind spots.
NetDeTox~\cite{NetDeTox} critiques GNN piracy detectors using a hybrid framework;
a reinforcement learning (RL) agent screens high-leverage netlist restructuring options,
and an LLM agent plans gate transformations, evading detection in 90\% of cases, while optimizing area.
%achieving negative area overhead.
TrojanGYM~\cite{TrojanGYM} implements an attack-defense agentic loop,
restructuring netlist properties until generated Trojans successfully evade GNN classifiers at an 83\% rate.

\subsubsection{Bug Detection and Code Analysis}
Common Weakness Enumerations (CWEs) often depend on microarchitectural context, making simple pattern-matching ineffective.
VeriCWEty~\cite{VeriCWEty} addresses this through a Verilog-tuned decoder that extracts module- and line-level embeddings;
paired with an ensemble-labeled dataset, the classifier achieves 89\% precision in identifying vulnerabilities and 96\% localization accuracy.
LASHED~\cite{LASHED} pairs static analysis with LLMs to contextualize errors and eliminate false positives,
while MARVEL~\cite{MARVEL} employs a hierarchical multi-agent setup (Linter, CWE, and RAG executors).
Finally, FLAG~\cite{FLAG} enables test-free fault localization by calculating token logprobs and semantic embedding distances directly from source code.

\subsubsection{Gap for Securing Chiplet Systems}
While LLMs have been used successfully for securing hardware in general, their dedicated application to securing 2.5D/3D systems is missing.
Current LLM setups lack semantic awareness of system-level interconnect fabrics, multi-vendor trust boundaries, and active interposer configurations.
Without domain-aware agents capable of synthesizing system-wide security constraints and mapping them to physical enforcement primitives, identifying hidden flaws in complex 2.5D/3D systems remains a severe 
bottleneck, leaving a window of vulnerabilities before human designers can deploy patches.
%Bridging this gap is imperative:
%unless intelligent agents are developed to synthesize system-wide security constraints and map them to physical enforcement primitives,
% attackers will inevitably find vulnerabilities in such complex 2.5D/3D systems faster than human designers can patch them.

\section{Conclusion and Outlook}

The convergence of 2.5D chiplet systems and LLM-accelerated EDA pipelines necessitates a dual paradigm shift toward system-level, secure-by-construction architectures on the one hand and
%Maximizing design utility and trustworthiness for LLM pipelines requires
orchestrated multi-agent loops operating within rigorously verified boundaries on the other hand.
%Layering an active interposer-based RoT defense
%with LLM-native security frameworks can mitigate cross-chiplet communication exploits, physical attacks, and EDA-specific supply chain vulnerabilities.
Looking beyond the significant advances in both domains, we call for a more synergetic ecosystem: deploying trusted, aligned LLM frameworks to verify and secure chiplet systems, while utilizing 2.5D-anchored RoT architectures to
accelerate and safeguard multi-vendor LLM deployment.
Accordingly, future research may utilize frontier models to bridge the gap between high-level security requirements and low-level physical enforcement mechanisms, e.g., for
automated generation of access policies for TRANSMONs and CMCs
%(for CMCs, APUs, DPUs, PRSs)
	directly from architectural descriptions and OS-initiated permissions.
%Such automation will enable self-calibrating runtime monitors that adaptively shield multi-vendor heterogeneous stacks from evolving microarchitectural threats.
Ultimately, closing the divide between LLM acceleration, EDA tooling, and physical integration is essential for all stakeholders
working on next-generation systems security.

%The convergence of 2.5D chiplet ecosystems and LLM-accelerated EDA pipelines exposes a critical cross-domain vulnerability: standard hardware security features ignore AI-driven design dynamics, while LLM-based security
%frameworks remain blind to the architectural realities of heterogeneous integration. This paper highlights this fundamental gap, demonstrating that while active interposers provide a robust physical RoT via TRANSMONs and
%CMCs, the security of the broader hardware lifecycle requires unified cross-domain orchestrators. Future research must aggressively target this intersection. Specifically, frontier reasoning models should be pioneered to
%dynamically bridge high-level security specifications with low-level interposer configurations—automatically generating CMC, APU, and DPU policy register spaces directly from natural-language architecture metrics.
%Closing the divide between AI automation and physical integration is the only viable path to rendering next-generation multi-vendor systems secure-by-construction.

%NOTE: Explore the use of LLMs to generate the CMC APU policies automatically from architectural specs.
%NOTE: LLMs can bridge the gap between high-level security requirements and the low-level PRSs used by the active interposer RoT.
%
%Looking beyond advances in these individual fields, we call for synergetic efforts
%like trusted LLM frameworks supporting the secure operation of chiplet systems on the one hand and
%trusted chiplet systems securing the deployment of LLM acceleration on the other hand.

\printbibliography

\end{document}